\documentstyle[12pt,epsfig,amsfonts]{article}

\catcode`\@=11
\def\@citex[#1]#2{%
\if@filesw \immediate \write \@auxout {\string \citation {#2}}\fi
\@tempcntb\m@ne \let\@h@ld\relax \def\@citea{}%
\@cite{%
  \@for \@citeb:=#2\do {%
    \@ifundefined {b@\@citeb}%
      {\@h@ld\@citea\@tempcntb\m@ne{\bf ?}%
      \@warning {Citation `\@citeb ' on page \thepage \space undefined}}%
      {\@tempcnta\@tempcntb \advance\@tempcnta\@ne%
      \@tempcntb\number\csname b@\@citeb \endcsname \relax%
      \ifnum\@tempcnta=\@tempcntb %Number follows previous--hold on to it
        \ifx\@h@ld\relax%
          \edef \@h@ld{\@citea\csname b@\@citeb\endcsname}%
        \else%
          \edef\@h@ld{\ifmmode{-}\else--\fi\csname b@\@citeb\endcsname}%
        \fi%
      \else%   %  non-successor--dump what's held and do this one
        \@h@ld\@citea\csname b@\@citeb \endcsname%
        \let\@h@ld\relax%
      \fi}%
    \def\@citea{,\penalty\@highpenalty\,}%
  }\@h@ld
}{#1}}

\def\@citeb#1#2{{[#1]\if@tempswa , #2\fi}}
\def\@citeu#1#2{{$^{#1}$\if@tempswa , #2\fi }}
\def\@citep#1#2{{#1\if@tempswa , #2\fi}}

\def\bcites{         % cite with []'s
        \catcode`\@=11
        \let\@cite=\@citeb
        \catcode`\@=12
}

\def\upcites{         % cite with exponents
        \catcode`\@=11
        \let\@cite=\@citeu
        \catcode`\@=12
}

\def\plaincites{      % cite without brackets
        \catcode`\@=11
        \let\@cite=\@citep
        \catcode`\@=12
}

\newcount\hour
\newcount\minute
\newtoks\amorpm
\hour=\time\divide\hour by 60
\minute=\time{\multiply\hour by 60 \global\advance\minute by-\hour}
\edef\standardtime{{\ifnum\hour<12 \global\amorpm={am}%
        \else\global\amorpm={pm}\advance\hour by-12 \fi
        \ifnum\hour=0 \hour=12 \fi
        \number\hour:\ifnum\minute<10 0\fi\number\minute\the\amorpm}}
\edef\militarytime{\number\hour:\ifnum\minute<10 0\fi\number\minute}

\def\draftlabel#1{{\@bsphack\if@filesw {\let\thepage\relax
   \xdef\@gtempa{\write\@auxout{\string
      \newlabel{#1}{{\@currentlabel}{\thepage}}}}}\@gtempa
   \if@nobreak \ifvmode\nobreak\fi\fi\fi\@esphack}
        \gdef\@eqnlabel{#1}}
\def\@eqnlabel{}
\def\@vacuum{}
\def\marginnote#1{}
\def\draftmarginnote#1{\marginpar{\raggedright\scriptsize\tt#1}}
\def\draft{
        \pagestyle{plain}
        \overfullrule=2pt
        \oddsidemargin -.5truein
        \def\@oddhead{\sl \phantom{\today\quad\militarytime} \hfil
        \smash{\Large\sl DRAFT} \hfil \today\quad\militarytime}
        \let\@evenhead\@oddhead
        \let\label=\draftlabel
        \let\marginnote=\draftmarginnote
        \def\ps@empty{\let\@mkboth\@gobbletwo
        \def\@oddfoot{\hfil \smash{\Large\sl DRAFT} \hfil}
        \let\@evenfoot\@oddhead}
        \def\@eqnnum{(\theequation)\rlap{\kern\marginparsep\tt\@eqnlabel}%
        \global\let\@eqnlabel\@vacuum}  }

\def\nblack{            % For people without blackboard fonts
        \def\ZZ{{Z \n{10} Z}}
        \def\NN{{N \n{14} N}}
        \def\CC{{C \n{11} C}}
        \def\RR{{R \n{11} R}}
        \def\QQ{{Q \n{12} Q}}
        \def\PP{{P \n{11} P}}
}

\def\eqalign#1{\null\,\vcenter{\openup\jot\m@th
  \ialign{\strut\hfil$\displaystyle{##}$&$\displaystyle{{}##}$\hfil
      \crcr#1\crcr}}\,}
\def\eqalignno#1{\displ@y \tabskip\centering
  \halign to\displaywidth{\hfil$\@lign\displaystyle{##}$\tabskip\z@skip
    &$\@lign\displaystyle{{}##}$\hfil\tabskip\centering
    &\llap{$\@lign##$}\tabskip\z@skip\crcr
    #1\crcr}}

\def\section{\@startsection {section}{1}{\z@}{3.ex plus 1ex minus
 .2ex}{2.ex plus .2ex}{\large\bf}}
\def\subsection{\@startsection{subsection}{2}{\z@}{2.75ex plus 1ex minus
 .2ex}{1.5ex plus .2ex}{\bf}}        

\def\appendix{{\newpage\section*{Appendix}}\let\appendix\section%
        {\setcounter{section}{0}
        \gdef\thesection{\Alph{section}}}\section}
\def\thefootnote{\arabic{footnote}}
\def\abstract{\if@twocolumn
\section*{Abstract}
\else %\small
\begin{center}
{\bf Abstract\vspace{-.5em}\vspace{0pt}}
\end{center}
\quotation
\fi}

\catcode`\@=12

\catcode`\@=11
\def\theequation{\arabic{equation}}
\catcode`\@=12

\nblack
\bcites

\nblack

\catcode`\@=11
\def\theequation{\thesection.\arabic{equation}}
\@addtoreset{equation}{section}
\@addtoreset{footnote}{section}
\@addtoreset{footnote}{subsection}
\catcode`\@=12

\newcommand{\beq}{\begin{equation}}
\newcommand{\beqa}{\begin{eqnarray}}
\newcommand{\bega}{\begin{array}}
\newcommand{\ea}{\end{array}}

\newcommand{\eeq}{\end{equation}}
\newcommand{\eeqa}{\end{eqnarray}}
\newcommand{\p}{\partial}

\newcommand{\Or}{{\cal O}}

\newcommand{\IR}{{\mathbb{R}}}

\begin{document}

\begin{titlepage}

\begin{center}
\hfill EFI-04-29 \\

\vskip 2.5 cm
{\large \bf Comments on D-brane Dynamics Near NS5-Branes}
\vskip 1 cm
\renewcommand{\thefootnote}{\fnsymbol{footnote}}
{
David A. Sahakyan%
\footnote{
        sahakian@theory.uchicago.edu
        }
}
\setcounter{footnote}{0}
\\
\vskip 0.5cm
{\sl Department of Physics, University of Chicago, \\

Chicago, IL 60637, USA\\ }

\end{center}

\vskip 0.5 cm
\begin{abstract}
We study the properties of a D-brane in the presence of $k$ NS5 branes.
The Dirac-Born-Infeld action describing the dynamics of this D-brane is
very similar to that of a non-BPS D-brane in ten dimensions.
As the D-brane approaches the fivebranes, its equation of state approaches that of a pressureless
fluid. In non-BPS D-brane case this is considered as an evidence for the decay of the
D-brane into ``tachyon matter''. We show that in our case similar behavior
is the consequence of the motion of the D-brane.
In particular in the rest frame of the moving D-brane the equation of state is that of a usual
D-brane, for which the pressure is equal to the energy density.  
We also compute the total cross-section for the decay of the D-brane into closed string modes and
show that the emitted energy has a power like divergence for $D0$, $D1$ and $D2$ branes, while converges for
higher dimensional D-branes. We also speculate on the possibility that the infalling D-brane
describes a decaying defect in six dimensional Little String Theory.

\end{abstract}
\end{titlepage}
\section{Introduction}
Recent years were marked by considerable progress in the understanding of
time-dependent physics in string theory.
One of the most studied subjects in this direction 
is the D-brane dynamics in the presence of an unstable (tachyonic) mode 
in the worldvolume theory. 
The condensation of this mode
leads to a new final time-independent state in the theory. 
For example in the case of non-BPS D-branes the condensation of the tachyonic mode 
may lead to a more stable brane configurations or to complete annihilation of the brane.
These systems can be studied by perturbing the  boundary CFT corresponding to the D-brane 
by exactly marginal operator corresponding to the tachyon 
\cite{Sen:2002nu,Sen:2002in,Sen:2002an,Sen:2002vv,Sen:2002qa,Gutperle:2002ai,Sugimoto:2002fp,Minahan:2002if,Hashimoto:2002xt,Mukhopadhyay:2002en,Rey:2003xs}. 
The recent development also showed that many properties of unstable D-branes are surprisingly well captured 
by Dirac-Born-Infeld (DBI) type actions 
\cite{Sen:1999md,Garousi:2000tr,Bergshoeff:2000dq,Kluson:2000iy,Kutasov:2003er,Niarchos:2004rw},
which turns out to be very useful tool for studying the D-brane decay.
In particular the DBI action analysis shows that the stress-energy tensor of the 
final state of the decaying D-brane is that of pressureless fluid.
This fact is considered as evidence for the decay of an unstable D-brane into pressureless fluid
of heavy closed string modes--``tachyon matter''.
This observation lead to the proposal of tree level open-closed string duality
\cite{Sen:2003xs,Sen:2003iv}, which states that the open string theory,
{\it without an explicit coupling to the closed strings} contains all the
necessary information about the closed string modes that the D-brane decays into.

Another example of a D-brane with an unstable mode in the worldvolume theory 
was recently considered in the paper by
D. Kutasov \cite{Kutasov:2004dj} (see also \cite{Yavartanoo:2004wb,Ghodsi:2004wn} for work along these lines). 
The setup discussed in this paper involved an (asymptotically) BPS $Dp$-brane, 
which is placed at some distance 
from a stack of $k$ parallel NS5-branes in type II
string theory. This configuration is not supersymmetric and hence
unstable--the D-brane will experience an attractive force, which
will move it towards the NS5 branes. 
This process is closely related to the tachyon condensation
problem. Indeed, as we will briefly review below, the DBI actions
describing the radial motion in the NS5 brane system and the tachyon decay for
non-BPS branes are related for large values of the tachyon field.
\footnote{It can be shown \cite{Kutasov:2004ct} that
the DBI actions are identical for all values of tachyon field if
one considers
a slightly more complicated system where one of the directions transverse to the
NS5-branes is compactified. }
Hence this system may serve as a useful model for studying the decay of non-BPS D-branes in 
ten dimensions. In particular the geometric interpretation of the tachyon direction may open 
new venues for understanding this problem better.

In this paper we make further progress in understanding 
the properties of a D-brane in the presence of NS5 branes.
In Section 2 we briefly review results of \cite{Kutasov:2004dj} and show that
the apparent decay of the D-brane into pressureless fluid, described by DBI action,
 is due to the fact 
that the D-brane moves with the speed close to the speed of light.
It turns out that in the rest frame of the D-brane the stress-energy tensor of it
is that of an ordinary D-brane with pressure equal to the energy density.
Hence we conclude that the DBI action describes only the classical
open string motion and the radiative corrections due to closed string emission 
should be accounted for separately.

In Section 3 we compute the closed string emission  and show that
the total energy of the closed strings
emitted is divergent for $Dp$-branes with $p\leq 2$, while finite
for $p>2$. This seems to suggest that the classical open string theory analysis is valid only
for $Dp$-branes with $p>2$, while for lower dimensional branes the large
radiative corrections render the classical trajectory invalid.

In Section 4 we speculate on the possibility that the infalling D-brane
describes holographically a decaying defect in the six dimensional 
Little String Theory (LST) \cite{Seiberg:1997zk}. We would like to argue that 
this defect is described by the same DBI action as the infalling D-brane and the
amplitude of the emission of little strings from the defect is \cite{LLM}  
\beq
I(\epsilon)=i\int dt \rho(t)e^{i\epsilon t}~,
\eeq
where $\rho(t)$ is related to the stress-energy tensor (pressure) as follows
\beq
T_{ij}\sim \rho(t)\delta_{ij}~.
\eeq
Then we show that the exponential part of full 
(non-perturbative) density of states of LST at high energies 
is such that it exactly cancels the exponentially divergent part 
of the total cross-section and 
hence this process is very reminiscent of the non-BPS D-brane decay.

In Section 5 we discuss our results and mention some directions for future work.

\section{Stress-Energy Tensor from Lorentz Transformation}
In  this section we 
show that the motion of D-branes described in
\cite{Kutasov:2004dj} using the DBI action 
does not include the backreaction from the radiation of closed strings. 
In particular we show that the stress-energy tensor, calculated in \cite{Kutasov:2004dj}
is that of a usual D-brane boosted to the speed close to the speed of light.

We start by reviewing the results of \cite{Kutasov:2004dj} on the radial motion of a $Dp$-brane in the background  
of $k$ NS5 branes. Let $x^\mu$, $\mu=0,1,\cdots, 6$ be the coordinates along the worldvolume of $NS5$-branes, 
while $x^n$, $n=6,7,8,9$ are labeling the transverse directions.
Then the background fields around $k$ $NS5$-branes are given by the CHS solution
\cite{Callan:1991at}. The metric, string coupling (dilaton) and NS-NS $B$ field are
\beq\label{CHS}
\bega{ll}
ds^2=dx^\mu dx_\mu+H(x^n)dx^mdx^m\equiv g_{MN}dx^M dx^N\\
{g_s(\Phi)\over g_s}=e^{(\Phi-\Phi_0)}=\sqrt{H(x^n)}\\
H_{mnp}=-\epsilon_{qmnp}\p^q\Phi~,
\ea
\eeq
where $H(x^n)$ is the harmonic function describing $k$ NS5 branes, $g_s$ is the asymptotic 
string coupling and $H_{mnp}$ is the field strength
of the $B$ field. We will be interested in the case of coincident fivebranes in which case $H(x^n)$
reduces to
\beq
H(r)=1+{kl_s^2\over r^2}~,
\eeq
with $r^2=x^n x_n$ and $l_s=\sqrt{\alpha^\prime}$.

Let us study the radial motion of a D-brane stretched in $(x^1,\cdots x^p)$
in this background. Without loss of generality we can label the worldvolume of the
D-brane by $(x^0,x^1,\cdots x^p)$. The position of the D-brane in the radial direction
gives rise to a scalar field $r(x^\mu)$ in the worldvolume theory.
The dynamics of this field is described by DBI action
\beq\label{DBI}
S_p=-\tau_p\int d^{p+1}x e^{-(\Phi-\Phi_0)}\sqrt{-det(G_{\mu\nu}+B_{\mu\nu})}~,
\eeq
where $\tau_p$ is the asymptotic tension of the $Dp$-brane
\beq
\tau_p\sim {1\over g_s l_s^{p+1}}~,
\eeq
$G_{\mu\nu}$ and $B_{\mu\nu}$ are the induced metric and the $B$ field respectively
\beq
\bega{ll}
G_{\mu\nu}={\p x^M\over \p x^\mu}{\p x^N\over \p x^\nu} g_{MN}\\
B_{\mu\nu}={\p x^M\over \p x^\mu}{\p x^N\over \p x^\nu} B_{MN}=0~.
\ea
\eeq
In the last line we used the fact that the only non-zero components of the $B$ field 
are in the angular, transverse  directions. 

In this paper we will discuss the spatially homogeneous
motion of the D-brane. Then $r$ is a function of $x^0=t$ only and using (\ref{DBI})
we find the following action describing the dynamics of this field
\beq
S_p=-\tau_p V\int dt\sqrt{{1\over H(r)}-\dot r^2}~,
\eeq
We will be mainly interested in the dynamics of the D-brane deep in the CHS throat, that is for 
$r<<\sqrt k l_s$. We would also like to take the asymptotic $g_s$ to zero keeping the rescaled 
radial coordinate $R\equiv g_s^{-1} r$ fixed.
This limit is usually taken in the holographic description of the Little String Theory \cite{Aharony:1998ub}.
 In this regime the $1$ in the harmonic function $H(r)$ can be neglected and the action 
takes the following form
\beq\label{action}
S_p=-T_p V\int dt {R\over \sqrt k l_s}\sqrt{1-\left({d\over dt}\log R\right)^2}~,
\eeq
where we introduced the rescaled D-brane tension $T_p\equiv\tau_p g_s$.
The action (\ref{action}) can be further simplified by introducing
\beq
e^{\phi\over \sqrt kl_s}\equiv{R\over \sqrt kl_s}~.
\eeq
In the new variables we find
\beq\label{DBINS}
S=-T_p V\int dt e^{\phi\over\sqrt k l_s}\sqrt{1-\dot\phi^2}~.
\eeq
This action is very reminiscent of the DBI action for the tachyon of  
the non-BPS D-brane (see {\it e.g.} \cite{Kutasov:2003er})
\beq\label{DBItach}
S\sim\int dt V(T)\sqrt{1-\dot T^2}~,
\eeq  
where 
\beq\label{poten}
V(T)={1\over\cosh{\alpha T\over 2l_s}}~.
\eeq
Here $\alpha$ is $1$ for bosonic string and $\sqrt 2$ for superstring.
We see that for large $T$ the map $\phi\rightarrow -T$, 
maps (\ref{DBINS}) into (\ref{DBItach}). Moreover for $k=2$ 
the action (\ref{DBINS}) is mimicking superstring, while for $k=4$ 
the bosonic string.

The equation of motion following from the action (\ref{DBINS}) is
\beq
1=-{\ddot\phi\sqrt k l_s\over 1-\dot\phi^2}~.
\eeq
It can be easily solved, yielding
\beq\label{eom}
e^{-{\phi\over\sqrt k l_s}}=A\cosh {t\over \sqrt k l_s}~.
\eeq
It is useful to express the constant $A$ in terms of the conserved energy of the brane
\beq
E=\dot\phi{\partial L\over \partial\dot\phi}-L=T_p V {e^{\phi\over\sqrt k l_s}\over\sqrt{1-\dot\phi^2}}~.
\eeq
Substituting this expression into (\ref{eom}) we find
\beq
e^{-{\phi\over\sqrt k l_s}}={T_p V\over E}\cosh {t\over \sqrt k l_s}~.
\eeq
In order to find the stress-energy tensor we should consider the full DBI action, involving
spatial excitations along the D-brane
\beq\label{DBIfull}
S_p=-T_p\int d^p x dt e^{\phi\over \sqrt k l_s}\sqrt{1-\dot\phi^2+(\p_i\phi)^2}\equiv \int {\cal L}dt d^p x~.
\eeq
Then using the definition
\beq
T_{\mu\nu}=\phi_{,\mu}{\p{\cal L}\over\p\phi_{,\nu}}+{\cal L}\eta_{\mu\nu}~,
\eeq
we find\footnote{It should be understood that the stress-energy is localized at the position of the 
D-brane, {\it i.e.} it involves delta functions in the directions transverse to the D-brane,
and in particular should have a factor of $\delta(\phi-\phi(t))$. The $T_{\mu\nu}$ computed from the
DBI action can be thought of as stress-energy tensor integrated over transverse directions.}
\beq\label{beftrn}
\bega{ll}
T_{00}=T_p {e^{\phi\over\sqrt k l_s}\over\sqrt{1-\dot\phi^2}}={E\over V}\\
T_{ij}=-T_p e^{\phi\over\sqrt k l_s}\sqrt{1-\dot\phi^2}\delta_{ij}~.
\ea
\eeq
The equations of motion imply that the $T_{00}$ is conserved, while the pressure 
is exponentially decaying
\beq\label{press}
T_{ij}V=-{E\over\cosh^2{t\over\sqrt k l_s}}~.
\eeq
A similar behavior of the pressure in the related problem of tachyon decay is
usually considered as evidence for the decay of an unstable D-brane into a pressureless fluid
of ``tachyon matter'' \cite{Sen:2003xs,Sen:2003iv}. We show below that here this behavior can be actually explained
from the fact that the D-brane is moving with the speed close to the speed of light.
Indeed consider the rest frame of the infalling D-brane at $t=t_0$
\beq\label{trans}
\bega{ll}
\tilde\phi={\phi-\phi(t_0)-v(t_0)(t-t_0)\over\sqrt{1-v^2(t_0)}}\\
\tilde t={t-t_0-v(t_0)(\phi-\phi(t_0))\over\sqrt{1-v^2(t_0)}}~,
\ea
\eeq
where $v(t_0)=\dot\phi(t_0)$. One can show that in this frame (at $t=t_0$)
the only non-zero components of
stress energy tensor take the form
\beq\label{serest}
\bega{ll}
\tilde T_{00}=M\delta(\tilde\phi)\\
\tilde T_{ij}=-M\delta_{ij}\delta(\tilde\phi)~,\\
%M=T_p e^{\phi(t_0)\over\sqrt{k}l_s}~,
\ea
\eeq
where $M=T_p e^{\phi(t_0)\over\sqrt{k}l_s}$.
Indeed, transforming (\ref{serest}) to the $(\phi,t)$ frame, we find
the following non-zero components of stress-energy tensor
\beq
\bega{ll}
T_{00}(t_0)={\left(\p\tilde t\over\p t\right)}^2 \tilde T_{00}={E\over V}\delta(\phi-\phi(t_0))\\
T_{ij}(t_0)=\tilde T_{00}=-{E\over V}{1\over\cosh^2{t_0\over\sqrt k l_s}}\delta(\phi-\phi(t_0))\\
T_{0\phi}(t_0)={\p\tilde t\over\p t}{\p\tilde t\over\p\phi}\tilde T_{00}=-{v(t_0)\over\sqrt{1-v^2(t_0)}}
T_p e^{\phi(t_0)\over\sqrt k l_s}\delta(\phi-\phi(t_0))~.
\ea
\eeq
The first two lines can be readily compared to (\ref{beftrn}), while the third line is the
energy density flow (momentum) in the $\phi$ direction, and can also be obtained from the DBI action
\footnote{The third line directly follows from energy conservation $\p^\nu T_{0\nu}=0$.}
\beq
T_{0\phi}=-{\delta{\cal L}\over\delta{\dot\phi}}~.
\eeq

\section{Emission rate from the one point function}
In this section we use worldsheet conformal field theory results to compute
the emission of closed strings from the infalling D-brane.
Our main interest here is to establish whether the radiative corrections spoil the
classical result obtained from the DBI analysis, {\it i.e.} whether the total emitted energy 
is divergent or not. Hence we will be somewhat cavalier
about the overall constants.

The string theory  deep in the CHS throat (\ref{CHS}) is described by an exact conformal field theory
\beq\label{back}
CFT=\IR^{5,1}\times SU(2)_k\times \IR_\phi~,
\eeq
The first factor in (\ref{back})
describes the space-time directions along the fivebranes.
The second factor describes the angular three-sphere.\footnote{As is well known, CFT on a 
three-sphere 
with a suitable NS $B_{\mu\nu}$ field is described by the $SU(2)$ 
WZW model.} The radius of the three-sphere is\footnote{In the rest of the paper we set $l_s=1$ for simplicity.}
\beq
R_{\rm sphere}=\sqrt{k}~.
\eeq
Finally, the third factor in (\ref{back}) describes the linear dilaton direction $\phi$
\beq
\Phi=-{Q\over 2}\phi;\qquad Q={2\over\sqrt k}~.
\eeq

The number of fivebranes $k$ determines the level of the 
$SU(2)$ current algebra and the slope of the dilaton in (\ref{back}). More precisely,
since (\ref{back}) is a background for the superstring, the worldsheet
theory contains fermions; the total level $k$ of the $SU(2)$ 
current algebra receives a contribution of $k-2$ 
from the bosons, and $+2$ from the fermions.
The total central charge is
\beq
6+{3(k-2)\over k}+\left(1+{3\over 2}Q^2\right)+10\cdot{1\over 2}=
6+{3(k-2)\over k}+\left(1+{6\over k}\right)+10\cdot{1\over 2}=15~,
\eeq
which is the correct value for the superstring.

The bosonic part of a normalizable closed string vertex operator in this background can be 
schematically represented as
\beq
V_b=P({\rm oscillators})\Phi_{j,m,\bar m} e^{Q\tilde j\phi}e^{i\vec k\cdot\vec X}e^{i\epsilon X^0},
\eeq
where $\Phi_{j,m,\bar m}$ is a primary of $SU(2)_{k-2}$  WZW, $\vec k$ is the momentum in $\IR^5$, 
$\epsilon$ is the energy and
\beq
\tilde j=-{1\over 2}+i\lambda,\quad \lambda\in \IR_+~,
\eeq
labels normalizable operators in the linear dilaton theory.
$P({\rm oscillators})$ is a polynomial in oscillators for $SU(2)$ WZW, $\IR^{5,1}$ and $\IR_\phi$.
 We will be interested in NS-NS and RR
sector closed string vertex operators, 
which can be obtained by coupling $V_b$ to fermions in the standard way.

The infalling D-brane is described by a boundary state $|B\rangle$ in this conformal field theory and
serves as a time-dependent source for closed string modes.
The emission of closed strings is described in terms of the one point function.
More precisely the amplitude for the emission of a closed string mode $V$ is \cite{LLM}
\beq
{\cal A}\sim {U(V)\over \sqrt \epsilon}={\langle V|B\rangle\over\sqrt \epsilon}~,
\eeq
where $\epsilon$ is the energy of the emitted mode. 
The boundary state $|B\rangle$ can be naturally factorized into
\beq\label{bstate}
|B\rangle=|B\rangle_{\IR^5}\times|B\rangle_{SU(2)}\times|B\rangle_{t,\phi}~.
\eeq
Let us discuss each of the factors in turn.
The first factor in $|B\rangle$ is especially simple and can be schematically presented as
\beq
\int d^{5-p}k\sum_{\psi\in {\cal H}_L} e^{i\phi(\psi)}|\psi\rangle_L |\psi\rangle_R |k\rangle~,
\eeq
where the sum goes over all left-right symmetric oscillators  states, which are unit normalized. 
The phase $e^{i\phi(\psi)}$ can in principle be determined, but since 
we will be interested in the absolute value square of the amplitude, this phase is irrelevant for us.

The second factor is the boundary state in $SU(2)_k$ WZW model.
In general there are $k+1$ different boundary states corresponding to BPS D-branes\cite{Apikian:1998py,Nepomechie:2001bu} 
(along with the $\bar D$-branes, 
which have an opposite sign in front of RR component)
\beq
|\tilde l\rangle={1\over\sqrt 2}\sum_{2j=0}^k \left({S^l_j\over \sqrt {S^0_j}} |j\rangle\rangle_{NS}+{S'^l_j\over \sqrt {S'^0_j}} |j\rangle\rangle_{R}\right)~,
\eeq
where  $l$ and $j$ are half-integers labeling the $SU(2)$ primaries,
\footnote{The primary $|j\rangle$ has $SU(2)$ spin $j$.} 
and $S^l_j$, $S'^l_j$ are the components of the modular transformation matrix.
$|j\rangle\rangle$ is the Ishibashi state \cite{Ishibashi:1988kg}
\beq
|j\rangle\rangle_{NS,R}=\sum_{\psi\in {\cal H}_{NS,R}} e^{i\phi(j,\psi)}|j,\psi\rangle_L |j,\psi\rangle_R,
\eeq
where the sum goes over all left-right symmetric  states constructed over current algebra primary $|j\rangle$.
As in the flat space case, the phase $e^{i\phi(j,\psi)}$ is not relevant for our purposes. 
We will also see that the particular form of $S^l_j$ affects only the overall normalization
of the total cross-section, thus we do not need to write down it explicitly.  
The D-brane that we are interested in, is the one that looks like a point on $S^3$, 
corresponds to the $\tilde l=0$ boundary state.

Finally let us discuss the last factor in (\ref{bstate}).
This boundary state describes the motion in the $(\phi,t)$ plane.
As discussed above the trajectory of the infalling D-brane
is given by
\beq\label{braneh}
e^{-{\phi\over \sqrt k}}=A\cosh {t\over \sqrt k }~.
\eeq
Performing Wick rotation $t\rightarrow iY$ we find
\beq\label{traj}
e^{-{\phi\over \sqrt k}}=A\cos {Y\over \sqrt k }~.
\eeq
This is the supersymmetric version of the
bosonic hairpin brane of \cite{LVZ} 
\beq\label{hairp}
{1\over 2}\tilde r \exp{\left(-{X\over \sqrt k}\right)}-\cos{Y\over\sqrt{k+2}}=0~,
\eeq
and was recently discussed in \cite{NST}.
The one point function of closed string vertex in the presence of a hairpin brane is
\beq\label{onepto}
U^\nu(P,Q)\sim {\Gamma(-i\sqrt k P)\Gamma(1-i{P\over\sqrt{k}})\over 
\Gamma ({1\over 2}+\nu+{\sqrt{k}\over 2}(Q-iP))\Gamma ({1\over 2}-\nu-{\sqrt{k}\over 2}(Q+iP))}~,
\eeq
where $P$ and $Q$ are momenta in the $\phi$ and $Y$ directions and $\nu=0,1/2$ for
NS-NS and RR sectors respectively. The one point function in the RR sector is obtained
from that of the NS-NS sector by 1/2-spectral flow.   

We would like to analytically continue the (\ref{onepto}) to find the closed string 
emission amplitude for the infalling D-brane. The naive analytical continuation
$Q\rightarrow i\epsilon$, $\nu\rightarrow i\nu$ gives
\beq\label{onept}
U^\nu(P,\epsilon)\sim {\Gamma(-i\sqrt k P)\Gamma(1-i{P\over\sqrt{k}})\over 
\Gamma ({1\over 2}+i\nu+i{\sqrt{k}\over 2}(\epsilon-P))\Gamma ({1\over 2}-i\nu-i{\sqrt{k}\over 2}(\epsilon+P))}~.
\eeq
Taking absolute value square, we find the cross-section for emission of a string mode
\beq
|U^\nu|^2\sim{1\over \sinh\pi P\sqrt k\sinh{\pi P\over \sqrt k}}(\cosh(\pi \epsilon\sqrt k+\nu)+\cosh\pi P\sqrt k).
\eeq
We see that the leading contribution comes from small $P$, and it diverges exponentially 
with the energy
\beq
|U|^2\sim e^{\pi \sqrt k \epsilon}.
\eeq
This answer is physically unacceptable. The reason for which 
the naive analytic continuation does not work is explained in \cite{NST} and
is essentially due to the fact that the $Y$ coordinate of the hairpin is effectively
compact, while the $\epsilon$ is unbounded.

On the other hand the one point function can be directly computed (at least semiclassically)
by integrating the wavefunction of the closed string mode over the profile of the brane.
In the NS-NS sector we find
\beq
U(P,\epsilon)=\int d\rho d\tau e^{2\rho} 
e^{i\tilde \epsilon\tau} e^{(-1+is)\rho}\delta(e^\rho\cosh\tau-{1\over 2}\tilde r)~,
\eeq
where
\beq
s=-\sqrt k P;\quad \tau={t\over \sqrt k};\quad \tilde \epsilon=\epsilon\sqrt k; \quad\rho={\phi\over \sqrt k}~,
\eeq
and the factor of $e^{2\rho}$ comes from the measure in the $\phi$ direction.
Note that the $\delta$-function used in the calculation differs from the one obtained from (\ref{braneh})
by a $\tau$ dependent factor. This normalization is natural, since as we will see below, in this case the
one point function will depend on $\tilde r$ only through a phase.
The physical quantities such as the absolute value square of the one point function should not depend on
$\tilde r$, since it can be shifted away by $\phi$ redefinition. 
The integral can be brought to the form
\beq
U=\left(\tilde r\over 2\right)^{is}\int_{-\infty}^{\infty}d\tau e^{i\tilde \epsilon\tau} (\cosh\tau)^{-1-is}~.
\eeq
Before performing the integral note that the main contribution to it comes from 
$|t|<\sqrt k$ region, which means that we are in the weak coupling region (at least if the initial energy of the
brane $E$ is large enough), 
hence our calculation is reliable.
Performing the integral we find the semiclassical expression for the one point function
\beq\label{brnh}
U=\left(\tilde r\right)^{is} 
{\Gamma({1\over 2}+{i\sqrt k\over 2}(\epsilon-P))\Gamma({1\over 2}-{i\sqrt k\over 2}(\epsilon+P))\over \Gamma(1-iP\sqrt k)}~.
\eeq
The exact answer is \cite{NST}
\beq
U=\left(\tilde r\right)^{is} 
{\Gamma({1\over 2}+{i\sqrt k\over 2}(\epsilon-P))\Gamma({1\over 2}-{i\sqrt k\over 2}(\epsilon+P))\over \Gamma(1-iP\sqrt k)}
\Gamma\left(1-{iP\over\sqrt k}\right)~.
\eeq
The one point function in the RR-sector can be easily obtained by $1/2$-spectral flow, which amounts to a
shift in energy $\epsilon\rightarrow\epsilon+{1\over\sqrt k} $. We will see that only the high energy
behavior of the amplitude is relevant for our purposes, which is the same for NS-NS and RR sectors, 
hence in the rest of this section we will restrict the
discussion to the NS-NS sector.

Now we can write down the full cross-section for the emission of a closed string,
which has momentum $P$ in the linear dilaton direction, 
$\vec k_\perp$ in the directions in $\IR^5$ transverse to the D-brane, 
labeled by $j\times j$ primary in $S^3$ and has total oscillator level $n$
\beq\label{cross}
|U(j,P,\epsilon(k_\perp,j,P,n))|^2\simeq
{\pi^2 \over k} {\sinh(\pi P\sqrt k)\over (\cosh\pi \sqrt k \epsilon+\cosh\pi\sqrt k P)\sinh{\pi P\over\sqrt k}}S_0^j~.
\eeq
The energy is determined from the physical state condition\footnote{We neglected $\Or(1)$ constant since 
at any rate we are interested in the high energy behavior.}
\beq\label{energyy}
\epsilon=\sqrt{P^2+k_\perp^2+4n+{j(j+1)\over k}}~.
\eeq
Let us compute the total average number and the energy of particles produced during the decay.
As mentioned above we would like to see whether either of these quantities blows up. 
From the form of the cross-section for the emission we conclude that the only divergence can
come from the high energy/large $P$ region.
In this region we can neglect the term $j(j+1)/k$ in $\epsilon$ and then the only $j$ dependence
of the cross-section comes from the factor $S_0^j$. 
Moreover we need to know only 
the asymptotic density of closed strings that the D-brane can decay into.
It is clear that the D-brane can emit only left-right symmetric closed string modes,
since the boundary state $|B\rangle$ describing the D-brane is left-right symmetric.

The asymptotic density of left-right symmetric states is
\footnote{Authors of \cite{NST} estimated the total cross-section using 
the full density of states, which seems to be inconsistent with the left-right symmetry of the boundary
state.} \cite{Kutasov:1990sv} 
\beq
d_n\sim n^{-{q\over 2}-{1\over 2}} e^{\pi\sqrt {4n}\sqrt{2k-1\over k}},
\eeq
where $2q=6$ is the number of non-compact spatial directions.
The average number of particles can be approximated by
\beq\label{nfull}
\bega{ll}
N\sim 
%{2\pi^2\over k}
\sum_j S_0^j&\int dP dl d^d k_\perp l^{-q}{1\over 2 \sqrt{P^2+l^2+k_\perp^2} }\times\\
&\exp(-\pi\sqrt k\sqrt{P^2+l^2+k_\perp^2}+\pi\sqrt k P-{\pi\over\sqrt k}P +\pi l \sqrt{2k-1\over k})~,
\ea
\eeq
where $l^2=4n$, $d$ is the number of directions in the $\IR^5$ transverse to the brane and we neglected some 
overall $k$-dependent factors in the integral. It is convenient to
express the integral in spherical coordinates
\beq\label{npart}
\bega{ll}
N\sim 
%k^{-{d-q+3\over 2}}
&\int d\theta d\phi  dr 
(\cos\theta)^{d-1}(\sin\theta)^{1-q}(\sin\phi)^{-q}r^{d-q}\times\\
 &\exp(-\pi r(1-\sin\theta(\cos\phi(1-{1\over k})+\sqrt {2k-1\over k^2}\sin\phi)))~,
\ea
 \eeq
where
\beq
\bega{ll}
r=\sqrt k\sqrt{P^2+l^2+k_\perp^2}=\sqrt k \epsilon~,\\
P\sqrt k= r\sin\theta\cos\phi~,\\
l\sqrt k= r\sin\theta\sin\phi~,\\
k_\perp\sqrt k=r\cos\theta~.\\
%\lambda=\sqrt {2k-1\over k^2}~.
\ea
\eeq
Introducing
\beq\bega{ll}
\cos\chi=1-{1\over k}~,\\
\sin\chi=\sqrt {2k-1\over k^2}~,
\ea
\eeq
we see that the expression in the exponent of (\ref{npart}) is manifestly
non-positive
\beq
 \exp(-\pi r(1-\sin\theta(\cos\phi(1-{1\over k})+\sqrt {2k-1\over k^2}\sin\phi)))=
 \exp(-\pi r(1-\sin\theta\cos(\phi-\chi)))~,
\eeq
and hence the integral can only have power-like divergences in $r$, which come from 
the region $\phi\sim \chi$ and $\theta\sim \pi/2$.

Performing the integral by steepest descent, we find
\beq
N\sim 
%k^{-{d-q+3\over 2}}\left({2k-1\over k^2}\right)^{-q}
\int dr r^{{d\over 2}-q-{1\over 2}}~.
\eeq
Similarly one can find the total emitted energy 
\beq
{\cal E}\sim 
%k^{-{d-q+4\over 2}}\left({2k-1\over k^2}\right)^{-q}
\int dr r^{{d\over 2}-q+{1\over 2}}~.
\eeq
We see that the emitted energy is finite for $Dp$-branes with $p>2$ and infinite
for $p\leq 2$.
 This result is identical to the result of similar calculation
in the case of rolling tachyon \cite{LLM}.
This seems to suggest that the tree level open string theory analysis is valid only
for $Dp$-branes with $p>2$, while for lower dimensional branes the large
radiative corrections render the classical trajectory invalid.
It is still plausible  that the decay process of the lower dimensional branes
is dominated by particles with energies (masses) comparable to the energy of the
original brane.
\footnote{Here we are referring to the $6d$ masses of the particles, as measured by the observer living on 
fivebranes.
The $10d$ mass of the particles emitted can actually be quite small (for large $k$); from $10d$ point of view the main 
contribution to the energy is coming from the 
momentum in the linear dilaton direction.}
%Finally we would like to find 
In this case the transverse momenta of the emitted particle with the energy $\epsilon$ are 
\beq\bega{ll}
\langle k_\perp\rangle=\epsilon\sqrt{2\over\pi r}\langle{\tilde\theta}\rangle\sim\sqrt {2\epsilon\over\pi\sqrt k}\\
\langle P\rangle\sim\epsilon (1-{1\over k})~.
\ea
\eeq
Hence we conclude that from the ten dimensional point of view the emitted particles 
are ultrarelativistic (at least for large $k$), 
that is they are moving with the speed close to the speed of light in the radial direction. On the other hand,
from the point of view of a six dimensional observer living on the fivebranes, 
these particles are highly non-relativistic.
We see that the six dimensional properties of the decay products in our case are again very 
reminiscent of the ten dimensional properties of the emitted particles in the
non-BPS D-brane decay.

\section{Holographic Description}
In the previous sections we saw that there is a striking similarity between
the decay of non-BPS D-brane in ten dimensions and the properties of (asymptotically)
BPS D-brane infalling onto a stack of NS5 branes. In particular we saw that
from the point of view of six-dimensional observer living on the fivebranes 
the decay products have exactly the 
same properties as the decay products of a non-BPS D-brane in ten dimensions.  
This leads us to the following proposal:{ \it The infalling BPS D-brane  describes holographically
a defect in the six dimensional Little String Theory, which resembles a non-BPS 
D-brane in type II string theory.}

In this section we attempt to  describe this defect directly in the LST
and thus provide further evidence for the proposal. 
To do that we will need to postulate few properties of defects in the LST, 
which make them very similar to non-BPS D-branes in ten dimensions. Using these properties
we will reproduce some of the results of section 3.
\begin{itemize}
\item[]{(1) The low energy behavior of a defect extended in $p+1$ space-time directions 
is described by the DBI action
(\ref{DBIfull}), in which the $\phi$ field is not regarded as a geometric direction but rather as 
a tachyon.}

\item[]{(2) Furthermore we require, following \cite{LLM}, 
that amplitude of the little string emission by the defect in LST
be given by
\beq
I(\epsilon)=i\int dt \rho(t)e^{i\epsilon t}~,
\eeq
where $\rho(t)$ is the (rescaled) pressure computed from the DBI action 
(\ref{press})
\beq
\rho(t)={1\over \cosh^2{t\over \sqrt k}}~.
\eeq}

\item[]{(3) The LST has a Hagedorn density of states \cite{Maldacena:1997cg} with the temperature
\beq
T_H={1\over 2\pi\sqrt k}~.
\eeq
Not all possible LST states can be emitted in the decay of a defect, but rather 
``left-right'' symmetric states,
\footnote{One might wonder why, while in section 3 only the 
perturbative density of states of ten dimensional theory entered the calculation, 
here we are insisting on using full Hagedorn spectrum of LST, which has many more states.
This is a standard situation in holographically related theories, 
{\it e.g.} in LST in order to compute the  high energy  density of states we need to know only
the perturbative spectrum in the background of NS5 branes.} 
{\it i.e.} only the square root of the total Hagedorn
spectrum will enter into the calculation of the total cross-section.\footnote{We should emphasize that this
last assumption is very speculative, since unlike in perturbative string theory,
we do not have any reason to believe that the states in Little String Theory can be factorized into 
the product of left and right movers.}}
\end{itemize}
Let us now compute, using these assumptions, the total emission of little strings from the
decaying defect.
First we should compute the Fourier transform of the pressure
\beq
\int {e^{i\epsilon t}\over\cosh^2 {t\over \sqrt k}}dt=2\sqrt k
\int {e^\tau e^{{i\epsilon\sqrt k\over 2}\tau}\over(1+e^\tau)^2}d\tau~.
\eeq
The integral can be computed by closing the contour at infinity and picking up the contributions from the poles
\beq\label{emis}
I(\epsilon)\sim {\epsilon\over \sinh{\pi \epsilon\sqrt k\over 2}}\sim \epsilon e^{-\pi \epsilon\sqrt k\over 2}~.
\eeq
Hence we see that the leading behavior of the cross-section for producing a state at energy  $\epsilon$ 
is
\beq
|I(\epsilon)|^2\sim \epsilon^2 e^{-\pi \epsilon\sqrt k}.
\eeq
Then the total energy emitted is
\beq
{\cal E}\sim\int dM d^{5-p}k_\perp |I(\epsilon)|^2\sqrt{\rho_H(M)}~,
\eeq
where $\rho_H(M)$ is the full Hagedorn density of states of LST 
\beq
\rho_H(M)\sim e^{2\pi \sqrt k M}~,
\eeq
and the integration goes over the mass of the little string modes $M$ and the momenta $k_\perp$
transverse to the defect. The square root of the Hagedorn 
density of states gets canceled by the exponential part of 
$|I(\epsilon)|^2$,
and one should look at the next to the leading behavior of $\rho_H(M)$ to actually compute the emitted energy.
For the second quantized Little String Theory this was computed in \cite{Kutasov:2000jp}, but since the
LST is strongly interacting theory this result is not directly applicable here. 
We conclude that, just as in section 3, the emitted energy has power like behavior,
although we were not able to compute this power. This result suggests that our assumptions 
about the properties of defects in LST outlined above are correct and the defects in six dimensional LST behave
in the same way as non-BPS D-branes in ten dimensions.

\section{Discussion}
In this note we studied the properties of an (asymptotic) BPS D-brane in type II string theory 
in the presence of $k$ parallel NS5 branes. The classical open string dynamics of the 
D-brane is described by DBI action.
As D-brane approaches to the NS5 branes its equation
of state approaches to that of a pressureless fluid. 
We showed that the apparent decay of the D-brane
into pressureless fluid is due to its motion. More precisely we found the equation of state
of the D-brane in its rest frame and it is that of a usual brane
with the pressure equal to the energy density. This indicates that the DBI action
does not describe the closed string emission and the radiative corrections should be accounted for
separately. Next we computed the closed string emission, using exact CFT results,
and found that the emitted energy is finite for $Dp$-branes with $p>2$, while has power divergence
for the lower dimensional branes. Hence we concluded that the classical open string theory analysis is not applicable
for the $D0$, $D1$ and $D2$ branes. This result is very reminiscent of the conclusion that one arrives to in
the case of the decay of non-BPS $Dp$-branes \cite{LLM}, 
namely the emitted energy is finite for $p>2$ and diverges for $p\leq 2$. 
This lead us to the following proposal:
{ \it The infalling BPS D-brane  describes holographically
 a defect in the six dimensional Little String Theory, which resembles  a non-BPS 
D-brane in type II string theory.} 
Next we attempted to describe the defect directly in Little String Theory to provide 
further evidence for our proposal. 
Under assumptions, which are very natural from the
non-BPS D-brane prospective,  we reproduced some of the exact CFT results. Namely
we showed that the emitted energy in LST has power like behavior, thus confirming that
the defects in six dimensional LST behave in the same way as non-BPS D-branes 
in ten dimensions.

In conclusion we would like to discuss possible directions for future work. 
It would be interesting to generalize the results of this paper to other
holographically related theories. For example one could consider the dynamics of a
$D1$ brane in the near horizon geometry of $N$ $D3$ branes.
In this case the two holographically related theories are the type IIB theory on $AdS_5\times S^5$
and ${\cal N}=4$ supersymmetric Yang Mills theory in four dimensions \cite{Maldacena:1997re}.
A $D1$-brane will be attracted to the $D3$-branes \cite{Panigrahi:2004qr} and it is plausible
that, just as in the system considered in this paper, it will describe a decaying one dimensional
defect in ${\cal N}=4$ SYM. It would be interesting to identify this object and check whether
the descriptions on two sides agree. 

A more ambitious program would be to describe holographically 
non-BPS D-branes in ten dimensions
as some kind of moving defects in a higher dimensional theory, thus making the tachyon direction
geometric. At the moment it is not clear whether such a description is possible.

\section*{Acknowledgements}
I am grateful to D. Kutasov and A. Parnachev for useful discussions.
This work is supported in part by DOE grant \#DE-FG02-90ER40560.
%%%%%%%%%%%%%%%%%%%%%%%%%%%%%%%%%%%%%%%%%%%%%%%%%%%%%%%%%%%%%%%%%%%%

\end{document}